\documentclass[a4paper,twoside]{article}
\newcommand{\affil}[1]{$^{\rm #1}$}
\newcommand{\ms}{$M_{\odot}$}
\newcommand{\msb}{$M_{\odot}$~}
\newcommand{\ct}{$^{13}$C}

\usepackage[authoryear]{natbib}
\bibpunct{(}{)}{;}{a}{}{,}
\usepackage{graphicx}
\date{} %Please leave the date blank
\title{\large\bf\flushleft Nucleosynthesis of light element isotopes \\
in evolved stars experiencing extended mixing}
\author{\parbox{\textwidth}{\flushleft
\vspace{-0.5cm}
{\it S. Palmerini\affil{A,B,C}, M. Busso\affil{A,B}, E. Maiorca\affil{A,B} and R. Guandalini\affil{A,B}}\\
\vspace{0.4cm}
{\small \affil{A}\,Dipartimento di Fisica, Universit$\rm\grave{a}$ degli Studi di Perugia, via Pascoli, 06123 Perugia, Italy}\\
{\small \affil{B}\,INFN Sezione di Perugia, via Pascoli, 06123 Perugia, Italy}\\
{\small \affil{C}\,Email: palmerini@fisica.unipg.it}}}
\begin{document}
\twocolumn[
\begin{changemargin}{.8cm}{.5cm}
\begin{minipage}{.9\textwidth}
\vspace{-1cm}
\maketitle
%
%
%%%%%%%%%%%%%     ABSTRACT    %%%%%%%%%%%%%
%Abstract of no more than 200 words here.
\small{\bf Abstract: We present computations of nucleosynthesis in
red giants and asymptotic giant branch stars of Population I
experiencing extended mixing. The assumed physical cause for mass
transport is the buoyancy of magnetized structures, according to
recent suggestions. The peculiar property of such a mechanism is
to allow for both fast and slow mixing phenomena, as required for
reproducing the spread in Li abundances displayed by red giants
and as discussed in an accompanying paper. We explore here the
effects of this kind of mass transport on CNO and
intermediate-mass nuclei and compare the results with the
available evidence from evolved red giants and from the isotopic
composition of presolar grains of AGB origin. It is found that a
good general accord exists between predictions and measurements;
in this framework we also show which type of observational data
best constrains the various parameters. We conclude that magnetic
buoyancy, allowing for mixing at rather different speeds, can be
an interesting scenario to explore for explaining together the
abundances of CNO nuclei and of Li.}
%%%%%%%%%%%%%     KEYWORDS    %%%%%%%%%%%%%

\noindent


% Write keywords here
% Please write all keywords in lower case. PASA uses the
% standard list of subject headings adopted by The Astrophysical Journal
% and available from http://www.journals.uchicago.edu/ApJ/keywords_text.html.
% Keywords are separated by em-dashes, i.e. ---

%%%%%%%%DO NOT EDIT%%%%%%%%%%%%
\medskip
\medskip
\end{minipage}
\end{changemargin}
]
\small
%%%%%%%%EDIT FROM HERE%%%%%%%%%%%%

\section{Introduction}

The atmospheric composition of evolved red giants is known to result
from the interplay of nuclear and dynamical processes, the latter
transporting material from the burning shells to the envelope \citep{b99};
these phenomena are also recognized to depend strongly on the
stellar metallicity \citep{str1}. The best known outcome of the above
interplay is the appearance in the photosphere, during the Asymptotic
Giant Branch (AGB), of heavy nuclei generated by slow neutron captures
in the so-called s-process \citep{kbw89}. The main neutron source for
the neutrons was identified in
($\alpha$,n) reactions on \ct, occurring in the radiative He-rich layers
\citep{gal1}.

Other chemical peculiarities accompany the advan\-ced evolution of
low mass stars (LMS). They are only partly observed by stellar
spectroscopy \citep{har,sl90,wal}, important constraints on them
deriving also from high accuracy measurements of isotopic shifts
in presolar grains of AGB origin \citep{ama1,choi,nit1,zin1}. This
includes e.g. the ratios $^{12}$C/$^{13}$C, $^{17}$O/$^{16}$O,
$^{18}$O/$^{16}$O, $^{26}$Al/$^{27}$Al.

After the works by \citet{g89} and \citet{gb91}, it became clear that
canonical stellar models (including only purely convective mixing)
are unable to account for the whole set of changes affecting the isotopic
mix of light and intermediate-mass nuclei, which begin to appear during
the first ascent of the Red Giant Branch (RGB). Some additional transport
mechanisms is required \citep{bsw94,charb98,charb00,nol}, in which part of the
envelope material is carried down to regions of sufficiently high temperature to
undergo proton captures.

Such mixing phenomena were studied in several works
\citep{char94,wbs95,sb99} and called in various ways: extended
mixing, deep mixing, extra-mixing and cool bottom processes (or
CBP). They were often attributed to rotationally-induced effects
\citep{char94,charb98}; however, it was subsequently demonstrated
that the reaction of the stellar structure to a centrifugal
distortion would be too fast to allow for extended transport of
matter on long time scales \citep{pal}.

More recently, it was noticed by \cite{edl06} that important mixing processes can be induced in stars
by the molecular weight inversion generated in the reaction $^3{\rm He}+^3{\rm He} \rightarrow ^4{\rm He} + 2p$,
implementing a form of {\it thermohaline} diffusion \citep{char07}.

Alternative (or complementary) mechanisms exist: it was in
particular underlined that the presence of a magnetic dynamo would
permit the buoyancy of magnetized structures created near the
H-burning shell, thus inducing a form of matter circulation in RGB
and AGB stages \citep{bwnc,nor08,den}. Recently, similar buoyancy
processes for the Sun have been shown to reproduce a number of
details of the matter upflows and downflows and other known
features of the solar dynamo \citep{lisb}.

One has also to notice that extra-mixing during the AGB stage has
to account for the presence, in presolar grains of circumstellar
origin, of nuclei (like $^{26}$Al) that are synthesized well below
the layers where $^3$He burns, so that in late evolutionary stages
thermohaline mixing alone might not suffice \citep{palm}. This
process, with its typical low velocity, should be dominant in the
long-lasting main sequence phases, where it might account for the
element changes occurring in the Sun and in similar stars
\citep{mr07}. On the other hand, the scheme of magnetic buoyancy
devised by \citet{bwnc} can be faster, hence more suitable for the
post-main-sequence evolutionary stages, where in fact several
classes of magnetically active stars do exist \citep{and88}.

In a paper presented in this volume \citep{guan} we analyzed the
advantages that magnetic buoyancy might offer for explaining the
complex phenomenology of Li production and destruction in red
giants. Hereinafter we shall refer to that work as to "paper I".
In the present work we want to derive the effects that the same
models would induce on CNO and other intermediate-mass nuclei, in
order to verify the compatibility of that idea with a wider set of
observational constraints. In Section 2 we discuss the models
adopted; in Sections 3 and 4 we present some comparisons with
abundance observations in stars and presolar grains, respectively.
Finally, in Section 5 preliminary conclusions are drawn.

\section{Buoyancy and Mixing} As mentioned, several models of the
coupled occurrence of nucleosynthesis and extended mixing were
presented in the past. In particular, \cite{nol} proposed a scheme
based on a conveyor-belt-like circulation for transporting matter
between the base of the convective envelope and the region
immediately above the H-burning shell, also showing that this
approach is equivalent to a diffusion treatment. The work
contained two main free parameters, namely the mixing rate
$\dot{M}$ and the maximum temperature  experienced by the
circulating material, $T_P$; its results suggested that the
observed isotopic shifts in presolar grains require $\dot M$
values in the broad range 10$^{-7}$ to 10$^{-5}$\ms$\rm yr^{-1}$
and $T_P$ values such that ($\log~T_P - \log~T_{\rm H}) \ge 0.1$,
where $T_{\rm H}$ is the temperature of the H-burning shell,
defined as the one at which the maximum energy is generated.

In a subsequent work by \citet{bwnc} it was suggested that the
driving force for mixing might be provided by the buoyancy of
magnetic flux  tubes formed near the H-burning shell. Those
authors derived, for the speed of this buoyancy, the relation:
$$
v(r)=\frac{1}{2}\cdot(\frac{\rho(r)}{\rho_{0}})^{\frac{3}{4}}(\frac{r}{r_{0}})^{-\frac{1}{4}}
(\frac{g_{0}a_{0}}{C_{D}})^{\frac{1}{2}}(\frac{B_{0}}{\sqrt{P(r)}})
\eqno(1)
$$
\cite[see][for details]{bwnc}. In the above equation $B_0$ is the
magnetic field somewhere near the H-burning shell (where $T =
T_P$), $a_0$ is the initial radius of flux tubes in the same
position, $C_D$ is the aerodynamic drag coefficient, and the other
quantities (pressure, temperature and density) are given by the
stellar structure, as computed with the FRANEC code. Since heat
exchanges between magnetized and non-magnetized matter are not considered
in this approach, the buoyancy is quite fast; it starts with an upward motion,
which also induces (by mass conservation) a downflow.
Fast matter transport driven by magnetic buoyancy is indeed observed
in the Sun \citep{brian}.

In the calculations of the present paper we shall assume that the
above prescriptions (included in what we shall define as "Model
A") be valid for the RGB phases immediately after the first
dredge-up. These stages witness the repeated crossing, by the
point representing a star, of the same zone in the HR diagram, so
that these points pile up and the luminosity function (LF)
displays a "bump". Stars at the LF-bump have just experienced the
advancement of the H-burning shell through the composition
discontinuity left behind by the first dredge-up, so that the
radiative region below the convective envelope has no gradient in
molecular weight, and any mixing phenomenon can occur easily. In
these stages, as early noticed by \cite{charb00}, Li is enhanced,
most probably by a mixing mechanism mimicking the so-called
Cameron-Fowler process, i.e. a transport of $^7$Be to the surface
at a speed fast enough to prevent it from experiencing proton or
electron captures along the path \citep{camfow}. In paper I we
showed that a mixing scheme like that envisaged by \cite {bwnc}
can in fact account for the observed Li production. With the
application of Model A we want here to infer the corresponding
effects on CNO and intermediate-mass nuclei.

It is however well known from solar observations that the
dynamical behaviour of photospheric structures (including those
generated by magnetic dynamos) is  complex in nature and displays
motions at very different velocities
\citep{toshifumi,svanda1,svanda2}. No observational information on
such processes in red giants, on their speed and on the fractional
area they might cover are within our reach. It was however
underlined recently \citep{den} that the buoyancy of large
structures in red giants should occur at a velocity lower than
estimated by \cite{bwnc}, because heat exchanges  with the
environment grow with the surface of the rising "bubbles". Slow
transport should favour Li destruction. Guided by this
consideration and by Li observations in RGB and AGB stars, in
paper I we assumed that, after the first fast phase, the buoyancy
slows down to an extra-mixing process similar to those previously
explored by \cite{wbs95} and by \cite{nol}, so that the Li
abundances decrease. In order to verify the effects on CNO
isotopes and other intermediate nuclei of those assumptions we,
too, explore this possibility: it will be referred to as "Model
B".

Concerning the actual speed of the mixing processes, in Model A we
assume for it a form like in equation (1), but with a suitable
scaling. The equation was derived by \cite{bwnc}, from considering
that an adequate recycling of the RGB envelope is obtained by
releasing about $\dot N \simeq 3\times 10^{-8}$ flux tubes per
second. If we want now to refer not to entire flux tubes, but to
{\it small} magnetized structures capable of moving at a fast rate
(e.g. parcels of magnetized material less massive than, say, one
tenth of a flux tube or so), then we need a higher release rate
(by a factor of ten in this case). We must then verify that the
rising velocity, although high, remains realistic, in particular
not larger than the Alfv\'{e}n speed. Let now $\hat v$ be the
average buoyancy velocity over the radiative layers, and let the
distance covered by the buoyancy be $\Delta r$; from the model by
\cite{bwnc} for an RGB star of 1.5 $M_{\odot}$, using their
equation (1):
$$
\dot N = {{\hat v}\over{\Delta r}}
\eqno(2)
$$
and assuming, as done in paper I, $\log T_{\rm H} - \log T_P$ =
0.3, one gets $\Delta r$ = 5$\times$10$^{10}$ cm and $\hat v$ =
100 m/sec. For the average density of the radiative layers below
the envelope in RGB phases, this value is lower than the
Alfv\'{e}n velocity ($B/\sqrt{4\pi \bar \rho}$) if B is larger
than about 10$^4$ G. This condition is satisfied even in the solar
tachocline and should therefore be easily met. We conclude that
Model A is not physically unreasonable.

For the second case (Model B) we assume, as made in \cite{nol},
that the circulation is slow, both along the ascending and the
descending path; for the sake of simplicity we assume that the two
flows occupy the same fractional surface  $f=0.5$ (at any distance
from the stellar centre). The velocity of the flow is in this case
set by:
$$
v(r) = {{\dot M} \over {4 \pi r^2 \rho f}}
\eqno(3)
\label{vd}
$$

For the range of $\dot M$ values discussed in paper I we obtain $v$
values (in cm/sec) ranging from a few tenths to a few.
The choices of $T_P$ are adopted from paper I, as they guarantee that
the observed Li abundances in RGB and AGB stars be reproduced.
We then solve the set of equations describing the abundance changes in the
mixed material, due to the nucleosynthesis and displacement processes. They
can be expressed through the total derivatives:
$$
\label{eq2}
\frac{dN_i}{dt}=\frac{\partial{N_i}}{\partial{t}}+\frac{\partial{N_i}}
{\partial{M}}\frac{\partial{M}}{\partial{R}}\frac{\partial{R}}{\partial{t}}
\eqno(4)
$$
where the partial time derivative is due to nucleosynthesis and the second term is due to mixing.
For the nuclear parameters we adopted the NACRE compilation \citep{NACRE}.
\begin{figure}[t!]
%\begin{center}
\includegraphics[width=\columnwidth]{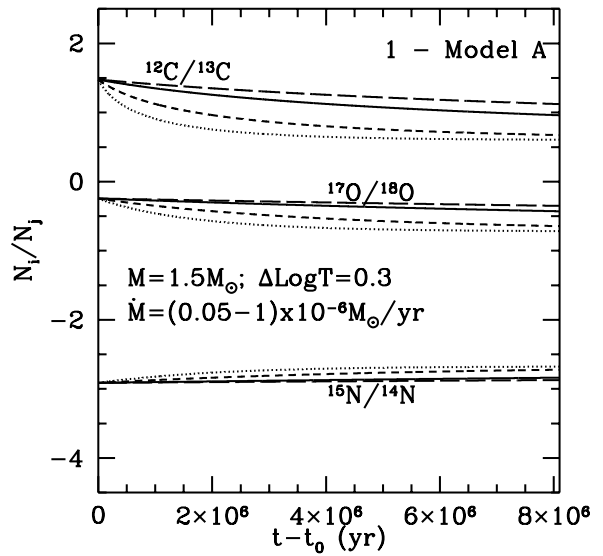}
\caption{The temporal evolution of CNO isotopic ratios in the
envelope of an RGB model with $M=1.5$ \msb and half solar
metallicity. The ratios derive from the operation of CBP,
following the prescription of Model A (see text for explanations).
Different lines refer to different choices for $\dot M$. Dots,
short dashes, continuous line and long dashes correspond to $\dot
M$ = 1, 0.6, 0.4 and 0.1 (in units of 10$^{-6}$\ms/yr),
respectively.}\label{rgb}
%\end{center}
\end{figure}

\begin{figure}[t!]
%\begin{center}
\includegraphics[width=\columnwidth]{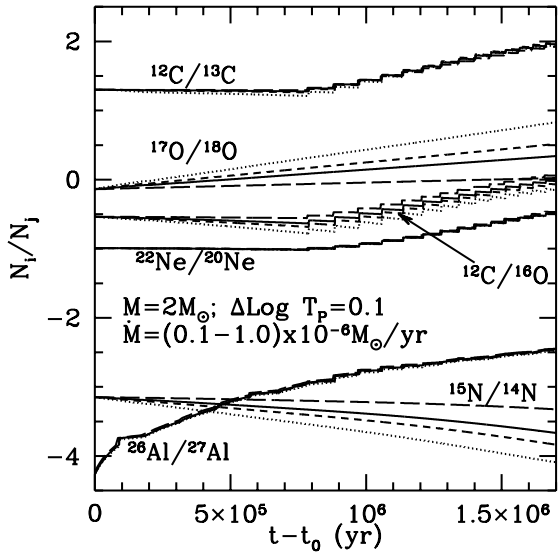}
\caption{The temporal evolution of isotopic ratios for CNO and Al in the envelope of an AGB model with $M = 2$ \msb and solar metallicity.
The ratios derive from the combined operation of CBP and TDU. The plot
shows the effects of changing the circulation rate $\dot M$ within a limited interval. Different lines refer to different choices for $\dot M$. Dots, short dashes, continuous line and long dashes correspond to
$\dot M$ = 1, 0.6, 0.4 and 0.1, respectively (in units of 10$^{-6}$\ms/yr).}.
\label{md}
%\end{center}
\end{figure}

\begin{figure}[t!]
%\begin{center}
\includegraphics[width=\columnwidth]{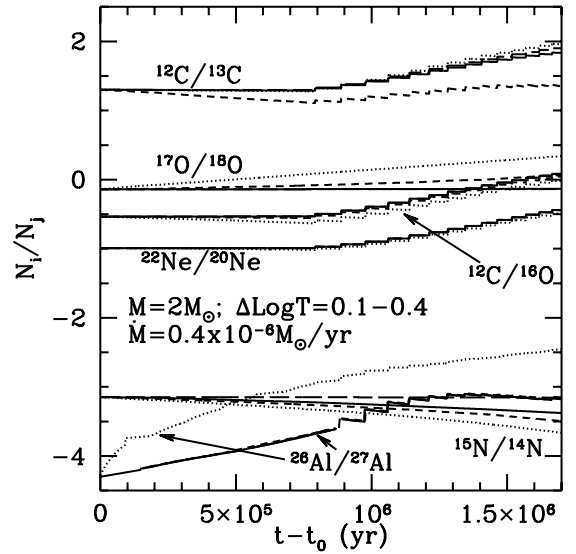}
\caption{Same as Figure 2, but showing the effects induced by changing the logarithmic separation $\Delta \log T$, between the H-shell temperature
$T_{\rm H}$ and the maximum temperature achieved by the circulating material $T_P$. Different lines refer to different choices for $\Delta \log T$. Dots, short dashes, continuous line and long dashes correspond to
$\Delta \log T$ = 0.1, 0.2, 0.3 and 0.4, respectively.}
\label{tp}
%\end{center}
\end{figure}

We study here AGB models of initially 1.5 \ms, at two different
metallicities ($Z = Z_{\odot}/2$,  and $Z_{\odot}/3$). We also
include the case of a 2 \msb star of solar metallicity.
Of the above models, the first one never becomes C-rich, and its predicted
composition can be compared  with those of O-rich AGB stars and of
presolar oxide grains \citep{choi}. The others do become carbon stars; their
predictions can be compared with the abundance observations in
C(N) giants and in presolar SiC grains \citep{zin1}.

The effects of extra-mixing in RGB and AGB stars are studied though a post-processing code that also reproduces
the effects of the third dredge up (TDU), occurring on the AGB after advanced {\it thermal pulses} of the He-burning shell.
The TDU contribution is important for several nuclei, in particular for $^{12}$C, $^{22}$Ne, $^{26}$Al, as well as for
neutron-capture nuclei produced in the He-intershell zone (which are, however, not considered here due to space
limitations).

It is well known that the efficiency of TDU decreases with increasing
metallicity \citep{bu01}. In our analysis the timing and mixed mass of
each dredge-up episode are calculated from the physical structure of the
original stellar models, using the analytical expressions suggested in
\cite{str}. The composition of the dredged-up material, in terms of the
light nuclei we study here, is recomputed on the basis of the
models presented by \citet{was06}.

The duration of the thermally-pulsing AGB stage is controlled by
mass loss; its efficiency is unfortunately an unknown parameter.
For illustration purposes we adopt here the simple parametrization
by Reimers (1975) fixing the free parameter $\eta$ to unity. With
this choice the final compositions of the models considered are
characterized by C/O ratios below 1 for the 1.5 \msb model with
$Z= Z_{\odot}/2$; of up to 2 for the 1.5 \ms, $Z= Z_{\odot}/3$
case; and of up to 1.5 for the 2.0 \ms, $Z= Z_{\odot}$ model. The
actual C/O value reached at the end of each calculation depends on
the effectiveness assumed for CBP, in particular on $\dot M$
(lower  $\dot M$ values allowing higher $^{12}$C abundances to be
reached).

\section{Results and Discussion} During the evolution, the mixing
events mentioned in the previous Section cause the envelope
composition to change gradually in time. Both elemental and
isotopic abundances are affected and the isotopic mix of each
element is followed in detail as the time proceeds. Our
computations produce sequences like those illustrated in Figures 1
to 3.

In particular, Figure 1 illustrates the effects on CNO isotopic
ratios of applying Model A, with circulation rates $\dot M$ in the
range (0.05-1)x10$^{-6}$ \ms/yr, until the bolometric magnitude
reaches about zero, exactly as  done in Paper I for reproducing
the Li abundances at the L-bump. In this way we can infer the
isotopic shifts for intermediate mass elements expected from the
same mixing scheme already shown to be suitable for explaining the
Li abundances. As shown by the Figure, a transport at high speed,
down to regions of moderately high temperature ($\Delta \log T =
0.3$), as implied by Model A, has detectable effects on few
isotopes, and mainly on the $^{12}$C/$^{13}$C ratio.  All the
cases shown in Figure 1 do produce Li at the level required by
observations, if the mixing is performed through the intermittent
release of magnetized bubbles traveling at high velocity. As the
carbon isotopic mix is modified differently by mixing with
different $\dot M$ rates, we expect that stars at the L-bump,
showing Li in their spectra with abundances up to $\log
\epsilon(Li)$ = 2 $-$ 2.5 (see Paper I) display a spread in carbon
isotopic ratios directly related to the $\dot M$ value. This
effect is weaker for N and O isotopes. All the curves of Figure 1
have, in their terminal part, a slope close to zero. This depends
on the fact that the mixing has been applied only down to regions
where $\Delta \log T = 0.3$, rather far from the H-burning shell,
where the nucleosynthesis effects are moderate. In these zones the
$^{12}$C/$^{13}$C ratio is not at its typical H-burning
equilibrium value (near 3.5); $^{12}$C is only marginally affected
by p-captures, producing some $^{13}$C. Since these effects are
rather small, during the 8 $\times$ 10$^6$ yr spent in mixing
according to Model A, the envelope abundances have time to reach
almost their asymptotic values. The subsequent switch to a slower
transport (model B), while destroying Li at various levels, does
not change remarkably the isotopic ratios, unless the maximum
temperature $T_P$ is increased.

The subsequent occurrence of extra-mixing on the AGB (followed
here through  the slow mixing scheme of Model B, as already done
in paper I) adds its effects to those accumulated in the previous
RGB stages, and interacts with the abundance changes induced in
advanced stages by the third dredge-up. For carbon, the effects of
TDU and CBP are opposite, and combine in a complex way. TDU adds
fresh $^{12}$C from the He shell, while CBP burns it through
p-captures. The result is a stepwise trend, as shown in Figures 2
and 3. The two Figures also illustrate the sensitivity of isotopic
ratios to the parameters of the mixing model. Indeed, varying
$\dot M$ (Figure 2) has effects  on N-, O- and C-isotopes, but
induces no change on Ne and Al. On the contrary, changing the
maximum temperature $T_P$ (Figure 3) mainly affects the Al
isotopic ratio: the representative curves drop below the bottom
axis of the graph in the cases with the lower values of $T_P$. A
more limited dependence on $T_P$ is shown also by the nitrogen and
oxygen isotopic ratios, while the $^{12}$C/$^{13}$C ratio is seen
to vary with $T_P$ only for the case ($\Delta \log T = 0.2$) that
samples the minimum of the $^{12}$C abundance in the radiative
layers.

\begin{figure}[t!]
%\begin{center}
\includegraphics[width=\columnwidth]{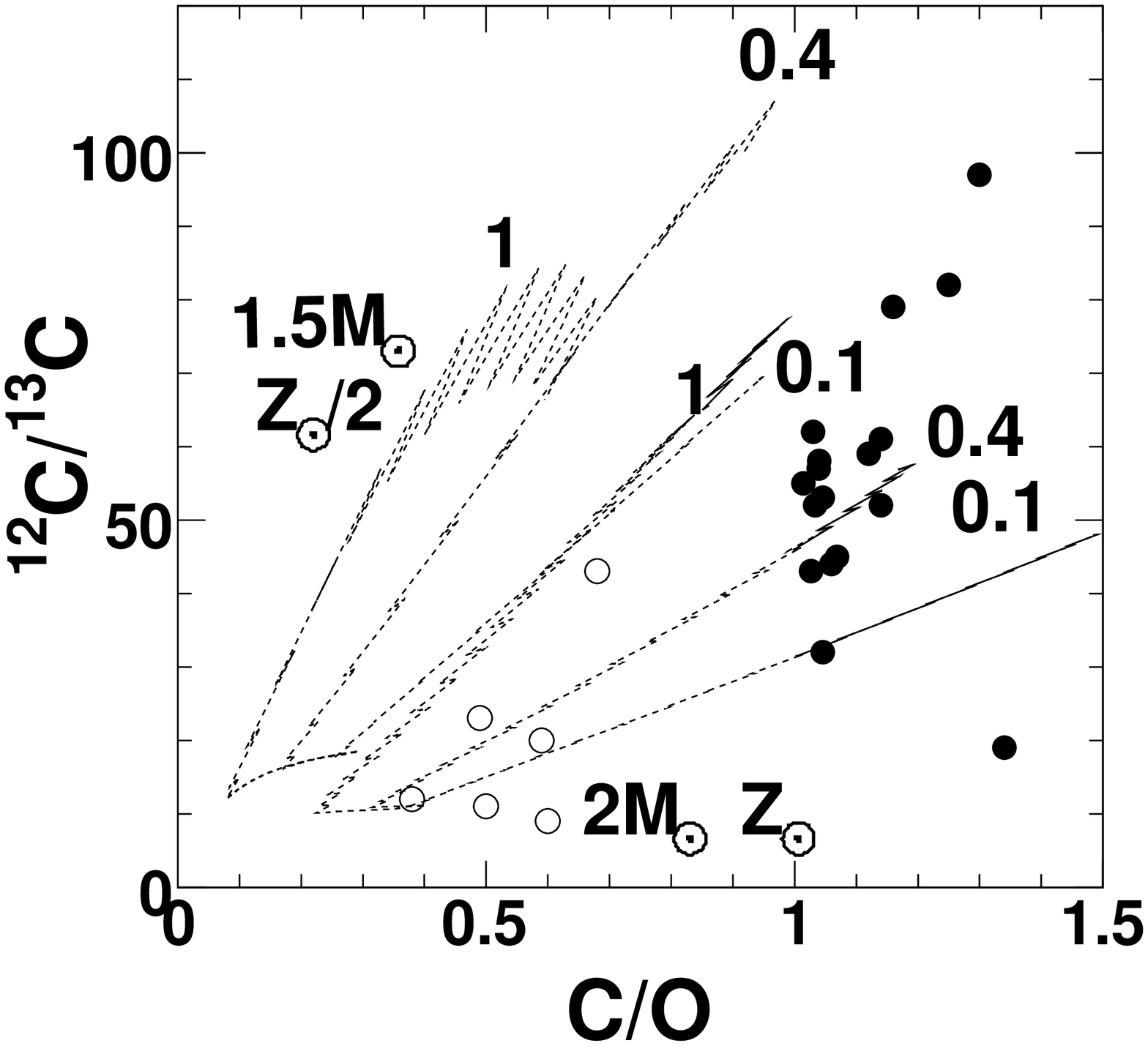}
\caption{A comparison of model predictions with observations for
the carbon isotope and C/O ratios, in red giants and AGB stars.
The sample of stars is the same for which, in paper I, with the
same mixing scheme and the same set of models, we reproduced the
Li abundances. Observations are from \cite{sl90,lam}. The lines
refer to the models of a 1.5 \msb star, with $Z=Z_{\odot}/3$ (the
set of curves extending at higher $^{12}$C/$^{13}$C ratios) and of
a 2 \msb star with $Z = Z_{\odot}$. Different values of $\dot M$
are indicated as labels, in units of 10$^{-6}$ \ms/yr. The
continuous portions of the curves indicate the phases where the
model envelopes are carbon-rich. Observational data are either
from M, MS, S giants (open circles) or from C(N) stars (Filled
circles).} \label{c12c13}
%\end{center}
\end{figure}

\begin{figure}[t!]
%\begin{center}
\includegraphics[width=\columnwidth]{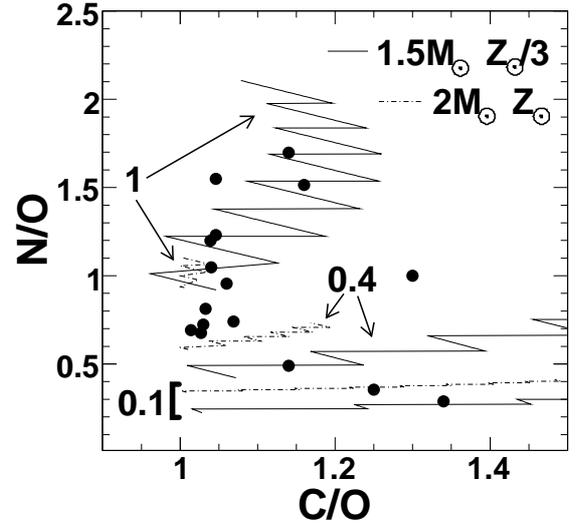}
\caption{A comparison of model predictions with observations for
the N/O and C/O ratios in carbon stars. The sample of C stars is
the same for which in paper I, with the same mixing scheme and the
same set of models, we reproduced the Li abundances. Observations
are from \cite{lam}. The lines refer to the models of a 1.5 \msb
star, with $Z=Z_{\odot}/3$ (continuous tracks) and of a 2 \msb
star with $Z = Z_{\odot}$ (dash-dotted tracks). Only the parts of
the model sequences corresponding to C/O ratios larger than unity
are shown. Different values of $\dot M$ are indicated as labels,
in units of 10$^{-6}$ \ms/yr.} \label{cono}
%\end{center}
\end{figure}

Figures 4 and 5 show some comparisons with observational data on
CNO abundances. The data points are from \cite{lam} and from
\cite{sl90}; they refer to stars for which our models were already
shown in paper I to account for the abundance of Li.  As is clear
from the plot, the curves that refer to the MS-S, O-rich, stages
and to the C(N), C-rich, ones pass correctly through the area of
the observed data. The curves have however a complex behaviour.
Figure 4 shows that, before TDU starts, pure extra-mixing above
the H-burning shell generates sequences that proceed leftward,
reducing the C/O and the carbon isotope ratio. When TDU begins to
operate this tendency is overcompensated by the addition, at each
mixing event, of rather large quantities of $^{12}$C. These are
subsequently partly burned by CBP in the interpulse periods. The
interplay of two opposite phenomena is seen also in Figure 5
(showing only models that achieve C/O ratios in excess of unity,
because the observed points are from C-stars; the case $M =
1.5$\ms, $Z = Z_{\odot}/2$ is not shown, as it never reaches the
C-star stage). At each dredge-up $^{12}$C increases (transported
from the He-rich layers) and also a small increase in N occurs
(transported from the inner H-shell layers, where N is very
abundant). In the subsequent interpulse stage CBP burns part of
the added carbon, further increasing N. The high N abundances
observed in many MS, S and C(N) giants are produced in this way;
the parent stars show only moderate Li enhancements, and this
certifies that they are of low mass. Indeed, the alternative
mechanism for N production, hot bottom burning in more massive
stars, would efficiently burn carbon (probably avoiding the
formation of a C star) and would produce high Li abundances, up to
$\log \epsilon(Li) = 4-5$.

\section{CNO isotopes in grains}

\begin{figure}
\begin{center}
\includegraphics[width=\columnwidth]{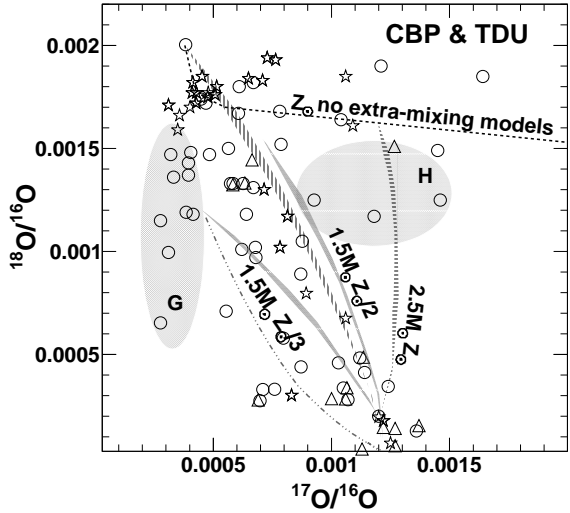}
\caption{The curves represent the model oxygen isotopic ratios in
the envelopes of population I AGB stars, experiencing extra-mixing
during the MS-S stage. The dots refer to presolar oxide grains
from the quoted references. The sequence labeled "no extra-mixing"
refers to the normal evolution of red giants as described by
models without CBP (the higher the stellar mass, the longer the
sequence extends toward the right). The other curves refer to our
models for different masses and metallicities. Shading along the
curves shows the effect of moderate dispersions in the values of
$\dot M$. Very high $\dot M$ values, much higher than considered
in paper I, would bend the curves toward the left. This is shown
by the dashed line, where $\dot M$ = 10$^{-5}$ \ms/yr. We include
also a special model, run for a 2.5 \msb star experiencing the
second dredge-up, with the aim of showing the maximum area that
CBP models can cover.}\label{fig6}
\end{center}
\end{figure}
Model results need also to be compared with the extensive set of
experimental data coming from presolar grains of
AGB origin. Examples of such comparisons are presented here as a preliminary check. They are a very small part
of what one wants to discuss, but for a question of space we must limit ourselves to a minimum.

A first important constraint to the models comes from Oxide
presolar grains, as shown in Figure 6. The Figure shows the
evolution followed by the oxygen isotopic mix in the RGB and AGB
atmosphere. Similarly to what was obtained by \citet{nol} with
parameterized calculations, also our new models appear to be well
suited to explain the composition of oxide grains that depart from
the canonical RGB-AGB sequence (represented by the data points and
by the dashed line in the upper part of the plot). The lines
representing our CBP models descend from the canonical one and
reach down to very low $^{18}$O/$^{16}$O ratios, covering the area
occupied by presolar grain measurements. Data in the G area can
only be fitted by models of low metallicity, while those in the H
area are from more massive AGB stars, experiencing the second
dredge-up and possibly hot bottom burning (we show the track of a
representative model doing the second dredge-up for illustrative
purposes).

\begin{figure}[h!!]
\includegraphics[width=\columnwidth]{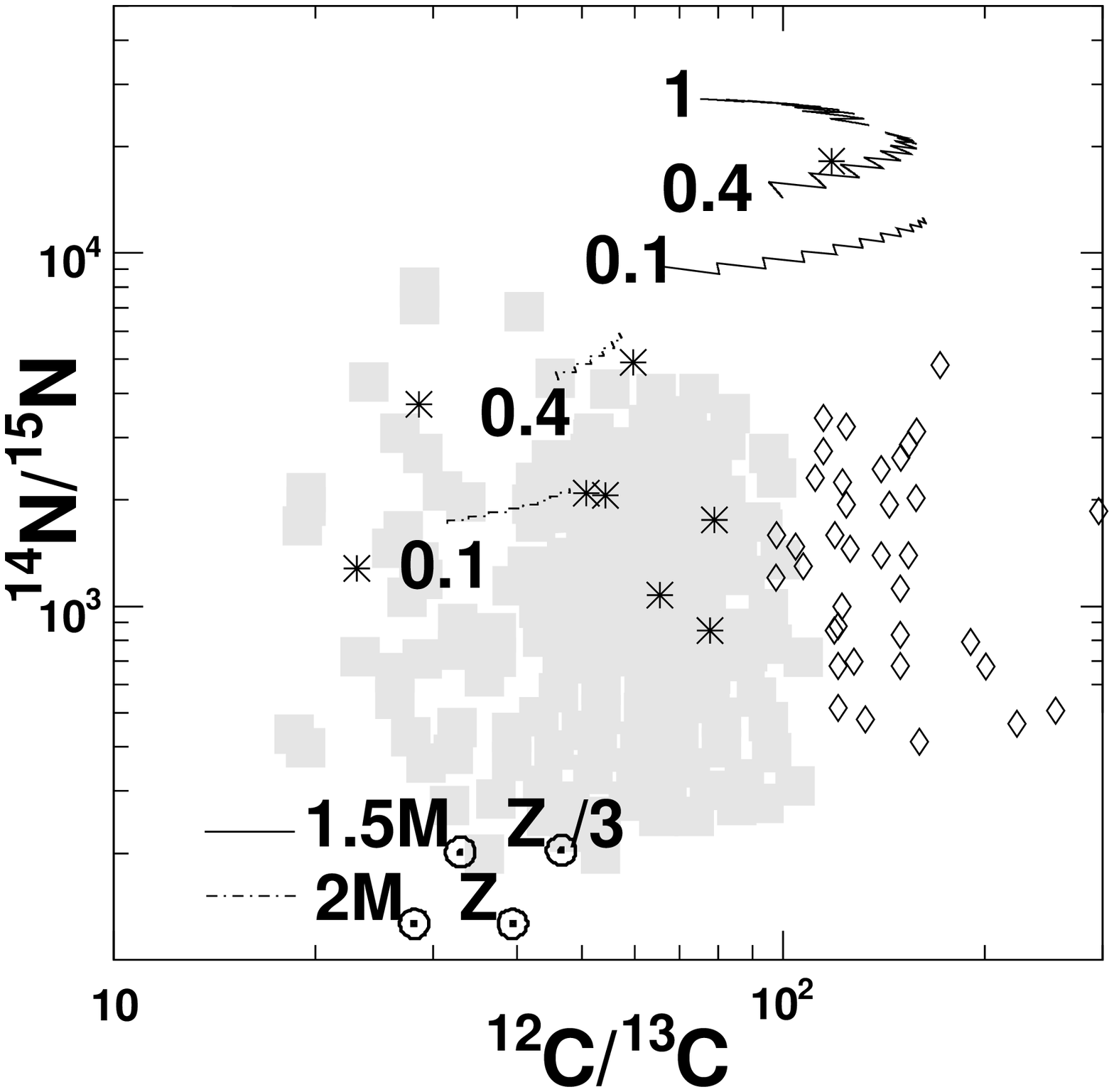}
\caption{The curves represent model sequences of $^{14}$N/$^{15}$N
versus $^{12}$C/$^{13}$C ratios in the envelopes of population I
C-rich AGB stars, experiencing extra-mixing. The $\dot M$ values
(in units of 10$^{-6}$\ms/yr) are shown as labels. The dots and
the grey zone refer to presolar SiC grains from lower-then-solar
and solar metallicty stars, respectively \citep{nit1,zin1}. Only
after a large downward shift, of at least a factor of 10, can the
models reproduce the data.
%This is in line with present doubts about contaminations of SiC grains with %$^{15}$N implanted from the solar wind.
}\label{fig7}
\end{figure}

The conclusions are rather obvious: i) population I AGB models can
account for the experimental data only if CBP is included; and ii)
in order to explain the grains with the lowest content of $^{18}$O
one needs to perform extensive extra-mixing. If this occurs at
high $\dot M$ values, then the curvature of model lines bends to
the left, as illustrated by the dash-dotted line, run with $\dot M
= 10^{-5}$ \ms/yr.

Figure 7 shows instead a comparison with measurements done in SiC
grains of AGB origin. The relevant models are those reaching
C-rich stages in the atmospheres. As some of our models yielding
C/O $>$ 1 are of lower-than-solar metallicity, we plot, for
comparison, the data not only for "mainstream" SiC grains (which
are believed to come from solar-metallicity stars), but also for
the Y and Z SiC grains, thought of as coming from C-rich envelopes
of slightly metal-poor AGB stars.

It is clear that, while for the carbon isotopic ratio solar
metallicity models well reproduce the mainstream grains, and lower
metallicity models occupy the correct region of Y and Z grains,
the measured nitrogen ratios are at odd with models by a large
amount. The discrepancy is at least by a factor of ten. We can
notice that direct comparisons of the models with nitrogen
abundances in stars (Figure 5) suggest that there is no conflict
between stellar observations and models (although for stars we do
not have the isotopic composition). The impossibility of fitting
the lower values of the $^{14}$N/$^{15}$N ratios in SiC grains is
a well known problem \citep{nol}. The most probable explanation is
that the grains undergo a contamination with solar $^{15}$N,
possibly implanted by the fast winds of the early Sun (J. Duprat,
private communication).

In none of the results for CNO nuclei, presented in this or in the
previous Section, through the application of our Model B and
through comparisons with either stellar observations or presolar
grain measurements, the physical origin of the mixing and the
velocity of transport play any relevant role. We adopted here the
same models used in paper I for explaining Li abundances, and we
can say that they are also suitable for understanding the
abundances of CNO nuclei in the same environments; however, only
from the need of producing Li we do derive a requirement on the
mixing velocity and hence an indication that magnetic buoyancy,
allowing for the fast upward transport of H-burning ashes, might
be at play, at least at the L-bump on the RGB.  Our preliminary
study then suggests that coordinated observations of Li and CNO
isotopes in the same stars are required to constrain the available
physical models for extra-mixing, disentangling their different
effects.

\section{Conclusions} In this paper we presented computations of
the temporal evolution of abundances at the surface of red giants
and AGB stars (especially for CNO and intermediate-mass nuclei),
as due to the occurrence of extra-mixing processes. We assumed two
mixing schemes, applied in sequence: one very fast, occurring at
and after the L-bump of the RGB, the second much slower and
occurring subsequently. This scenario was suggested by recent
works on magnetically-induced mixing in stars presenting a dynamo
mechanism \citep{bwnc,den} and is shown, in an accompanying paper,
to yield an explanation for the Li production and destruction in
RGB and AGB stars. The scope of our study was to derive the
consequences on other nuclei of the extra-mixing history inferred
from the Li constraint and to ascertain whether they agree with
the existing record of observations. The results of our (still
preliminary) exercise can be summarized as follows:

\begin{itemize}
\item It is in general possible to reproduce the Li and CNO
abundances observed in evolved red giants with models of
nucleosynthesis and extra-mixing occurring in the parent stars.

\item While explaining CNO isotopes alone does not yield special constraints on the mixing speed (but only on its effectiveness in terms of the transport rate $\dot M$ and of the maximum temperature achieved, $T_P$), including Li in the constraints informs us about the velocity with which the processed matter must travel trough the radiative layers below the convective envelope. This is so because for producing Li we need to save the parent nucleus $^7$Be to the envelope before it can undergo p- or e-captures along the path.

\item The joint reproduction of Li and CNO isotope abundances requires that extra-mixing be driven by a physical process capable of producing both fast and slow transport. The most common diffusive mechanisms, including thermohaline mixing, are not suitable for this task, while the circulation of magnetized matter driven by magnetic buoyancy can do the job.

\item Population I AGB stars of low mass are confirmed to be the
    site where many presolar oxygen-rich and carbon-rich grains form. The
    explanation of the isotopic mix of CNO elements displayed by the grains is a natural consequence of the mixing scheme presented here.

\item The above finding has an exception in the N isotopes of SiC presolar grains of AGB origin. Their excess in $^{15}$N cannot be reconciled with scenarios based on nucleosynthesis and mixing in low mass stars, and must come from a different source, possibly the injection of $^{15}$N in presolar material during the development of fast winds from the protosun.

\item The initial metallicity of the star has direct effects on internal mixing, through the efficiency of TDU, which increases for decreasing metal content. A more indirect effect occurs on CBP: it derives from the change of the abundances induced by TDU, over which proton captures
    occur in the interpulse phases.
\end{itemize}

\section*{Acknowledgments} We are indebted to the Italian Ministry
of Research for a PRIN grant (n.2006/022731) and to the National
Institute of Nuclear Physics (Section of Perugia, ERNA experiment)
for providing support and computing facilities.
%If needed

%\end{multicols}

\end{document}